\documentclass[amsmath,amssymb,aps,twocolumn,prb,showpacs]{revtex4}
\usepackage{graphicx}
\usepackage{dcolumn}
\usepackage{bm}
\usepackage{color}

\def\simge{\lower0.7ex\hbox{$\ \overset{>}{\sim}\ $}}
\def\simle{\lower0.7ex\hbox{$\ \overset{<}{\sim}\ $}}

\begin{document}

\title{Non-Fermi liquid, unscreened scalar chirality and parafermions in
a frustrated
tetrahedron Anderson model
}

\author{K. Hattori}
\author{H. Tsunetsugu}%
\affiliation{%
Institute for Solid State Physics, University of Tokyo, Kashiwanoha 5-1-5, Kashiwa, Chiba 277-8581, Japan
}%

\date{\today}

\begin{abstract}
We investigate a four-impurity Anderson model where localized 
 orbitals are located at vertices of a regular tetrahedron 
and find a novel fixed point in addition to the ordinary Fermi liquid
 phase. 
That is characterized by unscreened scalar chirality of a tetrahedron. 
In this phase,  parafermions emerges in the excitation spectrum and 
quasiparticle mass diverges as $1/|T\ln^3 T|$ at low
 temperatures ($T$). 
The diverging effective mass is a manifestation of singular Fermi liquid
 states 
as in the underscreened Kondo problem. 
Between the two phases, our Monte Carlo
 results show the existence of a 
non Fermi liquid critical point where the Kondo effects and the
 intersite antiferromagnetic interactions are valanced. Singular
 behaviors are prominent in the dynamics and we find that the frequency
 dependence of the self-energy is the marginal Fermi liquid like
$-{\rm Im}\Sigma\sim |\omega|$.
\end{abstract}

\maketitle

\section{Introduction}
The Kondo problem of multiple impurities has attracted much
attention since the pioneering work of the two-impurity system 
about 25 years ago.\cite{2imp} 
The presence of two competing energy scales, the Kondo temperature and
the intersite
exchange interaction, provides a nice play ground of quantum phase 
transition at zero temperature. Around the quantum critical point, non Fermi
liquid (NFL) behaviors emerge and 
thermodynamic quantities and their dynamics exhibit singular temperature 
dependence. 
These physics have a connection to quantum criticality in 
heavy fermion systems\cite{Lohn} and, thus, understanding much simpler cluster
problems is a key to capturing the nature of more complicated problems.

Quantum fluctuations in such strongly correlated systems 
can be controlled by tuning geometrical frustration. This is the case 
 not only in spin systems but also in itinerant electron systems and 
the frustration leads to novel ground states such as spin liquid
states.\cite{RVB}
As for the Kondo problem, frustration can be implemented by 
introducing a cluster of multiple impurities, and also controlling its
geometry. When the cluster is frustrated, 
 its ground state is degenerate due to nonmagnetic
degrees of freedom.
In the simplest case where the cluster is a triangle,\cite{exp} the ground state is degenerate in the
spin and the chirality sectors. Its Kondo problem was studied by the combination of 
 the numerical renormalization group (NRG) and boundary conformal field
theory (BCFT),\cite{3imp} and also by the continuous time quantum Monte Carlo
(CTQMC) method\cite{3imp2} and
 the renormalization group (RG).\cite{3imp3}
It has been established that the frustration 
 leads to a NFL phase. This phase is stable against 
perturbations 
that keep the triangular symmetry 
such as magnetic fields and particle-hole asymmetry,\cite{3imp} which
contrasts with the unstable NFL in the two-impurity case.\cite{2imp}

This paper reports extensive analysis on a minimal three-dimensional 
cluster problem with geometrical frustration,  
the Kondo problem of a regular tetrahedron. Tetrahedrons are simplex,
{\it i.e.}, the basic structure
for three dimensional ``frustrated'' materials such as
pyrochlore and spinel compounds. 
The result presented here would be
helpful for further studies on the lattice problem in addition to
the standpoint as a fundamental model analysis. 

This paper is organized as follows. In Sec. \ref{Mod}, we
introduce a microscopic model for four impurities.  We analyze
its two stable fixed points in Sec. \ref{2FP}, with emphasis on an 
unscreened phase. Section \ref{Monte} is 
devoted to clarifying the global phase diagram and discussions about a
critical point between the two stable fixed points. Finally, we 
summarize the present results in Sec. \ref{sum}


\section{Model} \label{Mod} We first introduce a four-impurity 
Anderson model, in which localized $s$-wave orbitals are located at
vertices of a regular tetrahedron and hybridize with a conduction 
band:
\begin{eqnarray}
{\mathcal H}&=&\sum_{\sigma}\int \frac{d{\bf k}}{(2\pi)^3}\epsilon_{\bf
 k}\psi^{\dagger}_{\bf k\sigma}\psi_{\bf k\sigma}+\sum_{i\sigma}
 \epsilon d_{i\sigma}^{\dagger}d_{i\sigma}\nonumber\\
&&+V_0\sum_{i\sigma} \int \frac{d{\bf k}}{\sqrt{(2\pi)^3}}(e^{i{\bf
 k}\cdot {\bf x}_i}d^{\dagger}_{i\sigma}\psi_{{\bf k}\sigma}+{\rm
 h.c.})\nonumber\\
&&
-t\sum_{i,j}d_{i\sigma}^{\dagger}d_{j\sigma}+U\sum_i
n_{i\uparrow}n_{i\downarrow},\label{H}
%
\end{eqnarray}
where $\psi^{\dagger}_{{\bf k}\sigma}$ creates a conduction 
electron with wave vector ${\bf k}$ and spin $\sigma$, $d_{i\sigma}^{\dagger}$
creates a localized electron at position ${\bf x}_i (i=1,2,3$ or $4)$,
$n_{i\sigma}=d^{\dagger}_{i\sigma}d_{i\sigma}$ and other parameters are conventional ones. We also introduce the
$d$-electron spin operator 
${\bf S}_i$ at ${\bf x}_i$, for later use. 

It is useful to transform $\psi$ to partial waves\cite{2imp} that couple with $d$
electrons. In the new basis, Hamiltonian (\ref{H}) is written as
\begin{eqnarray}
{\mathcal H}&=&\sum_{\alpha\sigma}\int dk\Big[\epsilon_{
 k}\psi^{\dagger}_{
 k\alpha\sigma}\psi_{k\alpha\sigma}+
V_{\alpha}(d^{\dagger}_{\alpha\sigma}\psi_{k\alpha\sigma}+{\rm h.c.})\Big]\nonumber\\
&&+\sum_{\alpha}\epsilon_{\alpha}d_{\alpha\sigma}^{\dagger}d_{\alpha\sigma}
+U {\rm \ terms}, \label{H2}
\end{eqnarray}
where the orbital index $\alpha$ runs over four channels: $(s,x,y,z)$. 
The orbital $s(x,y,z)$ is the basis of the $A_1$$(T_2)$ representation in
the $T_d$ point group.
The dependence of $V_{\alpha}$ on wave number $k$$=$$|\bf k|$ is 
ignored and replaced by the value at the Fermi wave number $k_F$: 
$V_s$$=$${V}[1+3\sin(k_Fa)/k_Fa]^{1/2}$ and
$V_{x,y,z}$$=$${V}[1-\sin(k_Fa)/k_Fa]^{1/2}$, with $a$$=$$|{\bf x}_{1}-{\bf x}_2|$ being
the impurity-impurity distance.
${V}$ is a parameter proportional to $V_0$.
The orbital energy is given as $\epsilon_{s}$$=$$\epsilon-3t$ and $\epsilon_{x,y,z}$$=$$\epsilon+t$.
 The new basis for $d$ electrons is given by
\begin{eqnarray}
d_{\alpha\sigma}=\frac{1}{2}\sum_{i}({\bf f}^{\alpha})_id_{i\sigma}, 
\end{eqnarray}
 with ${\bf f}^s$$=$$(1,1,1,1)$,
${\bf f}^x$$=$$(1,-1,1,-1)$, ${\bf f}^y$$=$$(1,-1,-1,1)$, and ${\bf f}^z$$=$$(1,1,-1,-1)$, 
and similar expressions for the conduction electrons.
 The bandwidth is set to $2D$ centered at the Fermi level 
and the density of states 
is set to constant $\rho=1/2D$
for all $\psi_{k\alpha\sigma}$.


\section{Two stable Fixed points}\label{2FP}
\begin{figure}[t]
\begin{center}
\includegraphics[width=0.5\textwidth]{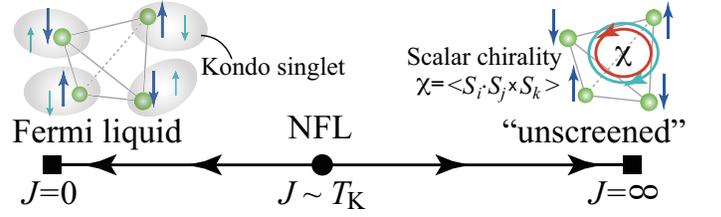}
\caption{\label{figRG} (Color online) Schematic renormalization flow
 in the Kondo regime as a function of intersite exchange
 interaction $J$. 
}
\end{center}
\end{figure}

First, we investigate two stable fixed points in the Kondo
regime, where the $d$ electron is nearly half filling  
$\sum_{\sigma}n_{i\sigma}$$\sim$$1$. Note that there are two
competing energy scales. One is the on-site Kondo energy
$T_K$$\sim$$D\exp(-|\epsilon|/4\rho V_0^2)$.
The other is the inter-site exchange interaction $J$, which
includes superexchange and Ruderman-Kittel-Kasuya-Yosida (RKKY) interactions. 
For $T_K$$\gg$$J$, the Fermi liquid (FL) fixed point is stable, where
each $d$ electron forms a spin singlet with conduction
electrons. In the opposite limit, $J$$\gg$$T_K$, four localized spins form
 spin singlets with a nonmagnetic double degeneracy: the $E$
 representation in the $T_d$ group. The two
states in the doublet can be characterized as eigenstates of the scalar
chirality (SC) 
\begin{eqnarray}
\chi={\bf S}_i\cdot ({\bf S}_j\times {\bf S}_k),
\end{eqnarray}
 with $i\ne
j\ne k$.\cite{Tsune}
The question is what happens when the interactions are switched on
between 
the chirality doublet and
conduction electrons?
We show that conduction electrons cannot 
 screen this doublet. 
Between the two stable fixed points, there exists a critical point as
shown in Fig. \ref{figRG}, and we discuss this in Sec. \ref{Monte}.

\subsection{Unscreened phase} 
\subsubsection{An effective  model}
In order to investigate the limit $J \gg T_K$,
we first derive an effective model of the local ground-state doublet. We
represent it 
by a pseudospin 1/2, whose bases are 
\begin{eqnarray}
|\!\Uparrow\rangle&=&\frac{1}{2}(1+\hat{I})( |
\uparrow\downarrow\uparrow\downarrow\rangle-|\uparrow\downarrow\downarrow\uparrow\rangle
),\\ 
|\!\Downarrow\rangle&=&\frac{1}{\sqrt{3}}(1+\hat{I})[|\uparrow\uparrow\downarrow\downarrow\rangle-\frac{1}{2}(|
\uparrow\downarrow\uparrow\downarrow\rangle+|\uparrow\downarrow\downarrow\uparrow\rangle)].
\end{eqnarray}
 Here, 
 $|\sigma_{1}\sigma_{2}\sigma_{3}\sigma_{4}\rangle$ represents a state with the
$z$ component of spin $\sigma_{i}$ at site $i$ and $\hat{I}$
is the spin inversion operator.
The effective model is strongly restricted by 
its symmetry. 
Since possible operators of the local $E$ doublet are 
$E$$\otimes$$E$$=$$A_1$$\oplus$$A_2$$\oplus$$E$, nontrivial couplings are those in the $A_2$ and $E$ sectors.
Taking into account the classifications for the conduction electron
bilinear operators, $T_2$$\otimes$$T_2$$=$$A_1$$\oplus$$E$$\oplus$$T_1$$\oplus$$T_2$,
$A_1$$\otimes$$A_1$$=$$A_1$, and $A_1$$\otimes$$T_2$$=$$T_2$, 
only a coupling in the $E$ sector is possible:
\begin{eqnarray}
{\mathcal H}_{\rm int} = g(\tau^z {\mathcal Q}+\tau^x{\mathcal T}),\label{HK}
\end{eqnarray}
where $g\propto t|V_x|^2/|\epsilon|^2$ for $U$$\to$$\infty$ and $(\tau^z,\tau^x)$
 are the spin-1/2 operators for the pseudospin with the $E$
 representation. The conduction electron parts,  
\begin{eqnarray}
\mathcal Q&\equiv&\frac{1}{\sqrt{6}}(\mathcal N_x+\mathcal N_y-2\mathcal N_z),\\
\mathcal T&\equiv&\frac{1}{\sqrt{2}}(\mathcal N_y-\mathcal N_x), 
\end{eqnarray}
 are the local orbital density with 
\begin{eqnarray}
\mathcal
N_{\alpha}\equiv\sum_{\sigma}\int
dkdk'\psi_{k\alpha\sigma}^{\dagger}\psi_{k'\alpha\sigma}.
\end{eqnarray}
They are the $E$ representation and the ``quadrupole'' operator with $[{\mathcal Q, \mathcal
T}]=0$. This 
commutativity distinguishes Eq. (\ref{HK}) from a conventional quadrupole Kondo
model\cite{quad} and leads novel behaviors as discussed below. 
Note that
${\mathcal N}_{s}$ does
not couple and the SC is represented by $\chi$$=$${\sqrt{3}}\tau^y$,
which is the $A_2$ representation and cannot appear in the Hamiltonian.


\subsubsection{Perturbative renormalization group}
Next, we analyze the effective Hamiltonian (\ref{HK}) by RG. 
It is remarkable that the one-loop order term vanishes and the leading order is  the two-loop level, and 
the RG equation is 
\begin{eqnarray}
D\frac{\partial
g}{\partial D}=\frac{1}{2}\rho^2g^3.
\end{eqnarray}
 The absence of
$O(g^2)$ terms in the RG equation is traced in $[\mathcal Q,\mathcal T]=0$. Solving
the RG equation, we obtain 
\begin{eqnarray}
\rho g_{\rm eff}=\frac{1}{\sqrt{\ln (T^*/D_{\rm eff})}}, 
\end{eqnarray}
 with
$T^*$$\equiv$$D\exp{[1/(\rho g)^{2}}]$ as the
effective bandwidth is decreased to 
$D_{\rm eff}$. This asymptotic logarithmic form is
similar to that in the underscreened Kondo model\cite{coleman} aside from the extra power 
1/2.  
Thus, the coupling constant approaches zero
with lowering temperature (energy). The impurity
degrees of freedom cannot be screened and there remains a residual
entropy $\ln 2$.
The pseudospin susceptibility has the Curie like temperature dependence
\begin{eqnarray}
\chi_{\rm zz}\equiv\int_0^{1/T}\langle
T_{\tau}\tau^z(\tau)\tau^z(0)\rangle d\tau
\sim T^{-1}, 
\end{eqnarray}
 where $T_{\tau}$
represents time-ordering.
Summing up the leading logarithmic terms, we find that 
the impurity specific heat coefficient $C_{\rm imp}/T$ diverges as
\begin{eqnarray}
\frac{C_{\rm imp}}{T}\sim \frac{1}{T\ln^3 (T^*/T)}. 
\end{eqnarray}
This means 
a diverging effective mass, and the power 3 is different from that in  ``singular''
Fermi liquid states.\cite{coleman}

\subsubsection{Boundary conformal field theory and numerical
   renormalization group analysis}
The presence of the unscreened phase is explicitly demonstrated in our
NRG\cite{NRG} calculations. Energy levels have degeneracy expected
for their quantum numbers (see below) plus an additional factor 2. 
This double degeneracy is a manifestation of the unscreened SC. 
The quantum numbers of each level consist of three parts: 
total charge $Q$, spin $j$, and the orbital part.
  The energies and the quantum numbers are in complete agreement
 with the prediction below derived via ``phase shift'' in the orbital sector in the BCFT 
language,\cite{CFT} assuming an unscreened SC. This
confirms that the unscreened local object in this phase is a SC, 
 or equivalently different configurations of two spin-singlet pairs.\cite{Tsune}

Let us explain our BCFT analysis. We first introduce a conformal
embedding suitable to Eq. 
(\ref{HK}). 
The free Hamiltonian is represented by in addition to the global U(1) charge
current, the spin and orbital currents following the Kac-Moody algebra 
SU(2)$_3$ and SU(3)$_2$, respectively. 
One possible conformal embedding is
U(1)$\otimes$SU(2)$_3\otimes$SU(3)$_2$, and the energy eigenvalues of
the free Hamiltonian are
represented as
\begin{eqnarray}
E^0=\frac{\pi v_F}{l}\Bigg[\frac{Q^2}{12}+\frac{j(j+1)}{5}+\frac{c_{\rm SU(3)}}{5}+
 {\rm integer}\Bigg], \label{emb1}
\end{eqnarray}
where we have defined the effective one-dimensional space in $[-l,l]$, $v_F$ is the
Fermi velocity, and $c_{\rm SU(3)}$ is the eigenvalue of the Casimir operator in the
SU(3) sector: $c_{\rm SU(3)}=0$ for ${\bf 1}$, 
$c_{\rm SU(3)}=4/3$ for ${\bf 3}$ and ${\bar{\bf 3}}$, 
$c_{\rm SU(3)}=10/3$ for ${\bf 6}$ and $\bar{\bf 6}$, and  
$c_{\rm SU(3)}=3$ for ${\bf 8}$. 

The interaction (\ref{HK}) preserves each of the charges in
 the $T_2$ orbital separately, although it breaks the orbital SU(3)
symmetry. This means 
there remains two conserved U(1) charges in the orbital sector:
\begin{eqnarray}
\bar{\mathcal Q}&\equiv&\frac{1}{\sqrt{6}}(\mathcal {\bar N}_x +\mathcal {\bar N}_y- 
2\mathcal{\bar N}_z),\\
\bar{\mathcal T}&\equiv& \frac{1}{\sqrt{2}}(\mathcal {\bar N}_y
-\mathcal {\bar N}_x),
\end{eqnarray}
 with 
\begin{eqnarray}
\mathcal{\bar N}_{\alpha}\equiv\sum_{\sigma}\int
dk\psi_{k\alpha\sigma}^{\dagger}\psi_{k\alpha\sigma}.
\end{eqnarray}
 Note that the
total orbital charge 
$\bar {\mathcal
N}_{\alpha}$ differs from the local density ${\mathcal N}_{\alpha}$ and
$Q=\sum_{\alpha}\bar{\mathcal N}_{\alpha}$.
Factorized out with these two U(1) charges, the CFT for the remaining orbital
part SU(3)$_2$/[U(1)]$^2$ is known as the parafermion (PF) one 
 with the central charge
$c=6/5$.\cite{Z3}
Thus, the energy spectra for the free Hamiltonian 
 are written as 
\begin{eqnarray}
E^0=\frac{\pi v_F}{l}\Bigg[\frac{Q^2}{12}+\frac{j(j+1)}{5}+\frac{\bar{\mathcal
 R}^2}{4}+\Delta+ {\rm integer}\Bigg], \label{E02}
\end{eqnarray}
 where 
$\bar{\mathcal R}^2\equiv\bar{\mathcal T}^2+\bar{\mathcal
Q}^2$ and $\Delta$ is the dimension of the primary fields in the PF 
 sector (see the caption in Table \ref{tbl-1}). Equations (\ref{emb1})
 and (\ref{E02}) give exactly the same spectra. Note that
there is O(2) symmetry between $\bar{\mathcal Q}$ and $\bar{\mathcal T}$.
$Q$ and ${\bar{\mathcal R}}$ terms in Eq. (\ref{E02}) can be also
represented as
\begin{eqnarray}
\frac{Q^2}{12}+\frac{\bar{\mathcal
 R}^2}{4}=\frac{{\bar{\mathcal N}_x}^2+{\bar{\mathcal N}_y}^2+{\bar{\mathcal N}_z}^2}{5},
\end{eqnarray}
which is symmetric with respect to the orbital indices, as it should be. 

\begin{figure}[t]
\begin{center}
    \includegraphics[width=0.48\textwidth]{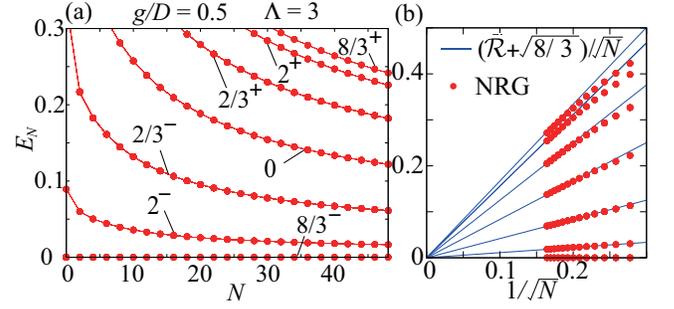}
\end{center}
\caption{(color online). (a) NRG spectra for odd $N$ for $g=0.5D$ and
 $\Lambda=3$. At each $N$, the energy of the lowest-energy states are
 subtracted. 
States are labeled by $\bar{\mathcal R}^2$ and the sign of the energy 
 listed in Table \ref{tbl-1}. (b)
 $E_{N}$ as a function of $1/\sqrt{N}.$
 }
\label{fig-NRG}
\end{figure}

Table \ref{List} and Fig.\ref{fig-NRG}(a)  show NRG spectra
obtained with keeping 10$^4$ states at each step. We have checked that the
results do not change when 10$^4$ more states are added. 
We find that the energy eigenvalues $E_N$ at the $N$th RG step
are given by the free
spectrum modified by an additional potential scattering in the O(2) sector:
\begin{eqnarray}
 E_N\propto \frac{l}{\pi v_F}E^0 \pm g_N \bar{\mathcal R},
\end{eqnarray}
 with $g_N$ constant. Note that this form remains
 O(2) symmetric and is derived from the free spectra by shifting $\bar{\mathcal R}\to \bar{\mathcal
R}\pm 2g_N$ in Eq. (\ref{E02}), {\it i.e.,} just a ``phase shift.'' The factor $\pm$ indicates
unscreened SC as in the Ising Kondo case.\cite{poor}
We note that the simple ``phase shift'' leads to PF excitation spectra for $g_N>0$.

The $N$th step in the NRG calculation is related to the
energy scale $D\Lambda^{-N/2}$ with
discretization parameter\cite{NRG} $\Lambda$ and we set
$\Lambda=3$. Replacing this for $D_{\rm eff}$ 
 in the two-loop expression of $g_{\rm eff}$, 
we find 
$g_N\propto 1/\sqrt{N}$. Indeed, we obtain
\begin{eqnarray}
E_N-E_{\infty}\propto\pm\frac{1}{\sqrt{N}}{\bar{\mathcal R}}, 
\end{eqnarray} 
in the NRG results as shown in Fig.\ref{fig-NRG} (b), which
confirms the two-loop results. This $1/\sqrt{N}$ dependence is a clear
contrast to $1/N$ dependence in underscreened and ferromagnetic Kondo models.

\begin{table}[t]
\caption{\label{List} Quantum numbers, energy, and
 degeneracy  of 128 low-energy states for odd
 $N$.  In 
the parafermion (PF) sector, following notations in Ref. \cite{Z3}, there are eight primary fields: \{$\mathbf 1,
 {\sigma}_{\uparrow},{\sigma}_{\downarrow},{\sigma}_3,{\psi}_1,{\psi}_2,{\psi}_{12},
 {\rho}$\}. ${\sigma}$'s and ${\psi}$'s
 are abbreviated simply as $\boldsymbol{\sigma}$ and
 $\boldsymbol{\psi}$, respectively. Their scaling dimension $\Delta$ is indicated
 in parentheses. The energy is measured from the value of the
 non-interacting ground states and then scaled appropriately. }
\begin{center}
\begin{tabular}{ccccccc}
\hline\hline
\vspace{-3mm}
\\
$\ \ Q\ \ $& $\ \ j\ \ \ \ $ &SU(3)&$\ \ \ \bar{{\mathcal R}}^2\
 \ \ $ & $\ \ {\rm PF}(\Delta)\ \ \ $ & Energy &degeneracy \vspace{0.5mm}\\
\hline
0&$3/2$&{\bf 1}&0&$\mathbf{1}(0)$&$0$&8\\
0&$1/2$&{\bf 8}&0& ${\rho}({3/5})$&$0$&4\\
&&{\bf }&0& ${\rho}({3/5})$&$0$&4\\
&&{\bf }&2& $\boldsymbol{\sigma}({1/10})$&$\pm \sqrt{2}$&12 each\\
$1$&$1$&{\bf 3}&${2/3}$& $\boldsymbol{\sigma}({1/10})$&$\pm
		     \sqrt{2/3}$&9 each\\
$1$&$0$&$\bar{\bf 6}$&${8/3}$& $\mathbf{1}(0)$&$\pm \sqrt{8/3}$&3 each\\
&&{\bf }&${2/3}$&$\boldsymbol{\psi}({1/2})$&$\pm \sqrt{2/3}$&3 each\\
$-1$&$1$&$\bar{\bf 3}$&${2/3}$&$\boldsymbol{\sigma}({1/10})$&$\pm
		     \sqrt{2/3}$&9 each\\
$-1$&$0$&{\bf 6}&$8/3$&$\mathbf{1}(0)$&$\pm \sqrt{8/3}$&3 each\\
&&{\bf }&$2/3$&$\boldsymbol{\psi}({1/2})$&$\pm \sqrt{2/3}$&3 each\\
2&$1/2$&$\bar{\bf 3}$&$2/3$&$\boldsymbol{\sigma}({1/10})$&$\pm
		     \sqrt{2/3}$&6 each\\
$-2$&$1/2$&{\bf 3}&$2/3$&$\boldsymbol{\sigma}({1/10})$&$\pm
		     \sqrt{2/3}$&6 each\\
$3$&$0$&{\bf 1}&$0$&$\mathbf{1}(0)$&$0$&2\\
$-3$&$0$&{\bf 1}&$0$&$\mathbf{1}(0)$&$0$&2\\
\hline
\hline
\end{tabular}
\end{center}
\label{tbl-1}
\end{table}

\subsubsection{A hidden angular momentum}
We have observed that the O(2) symmetry in the orbital space ${\mathcal
Q}$-${\mathcal T}$ is preserved in the energy spectrum of the
interacting system $g\ne 0$. 
This holds asymptotically where the bosonization becomes exact in the
low-energy limit. We demonstrate this below by explicit calculations.

The O(2) symmetry is related to a {\it hidden} conserved quantity of angular momentum. 
This angular momentum is nontrivial and defined in the two dimensional $\mathcal Q$-$\mathcal T$
space as  
\begin{eqnarray}
\mathcal L&\equiv& \sum_{n\ge 0}{\mathcal L}_n=
i\sum_{n>0} \frac{1}{n}[{\mathcal Q_n}\mathcal
T_{-n}-{\mathcal T}_n{\mathcal Q }_{-n}]+{\mathcal L}_0. 
\end{eqnarray}
Here, $\mathcal Q_n$ and $\mathcal T_n$ are the $n$th Fourier modes for
the U(1) bosons 
and 
\begin{eqnarray}
{\mathcal L}_0={\mathcal
Q}_0\theta_{\mathcal T}-{\mathcal T}_0\theta_{\mathcal Q},
\end{eqnarray}
 where
 $\theta_{\mathcal Q}$($\theta_{\mathcal T}$) is a canonical conjugate
 field to ${\mathcal Q}_0$(${\mathcal T}_0$), with $[{\mathcal Q}_0,\theta_{\mathcal Q}]\!=\![{\mathcal T}_0,\theta_{\mathcal T}]\!=\!i$. 
$\mathcal L$ is conserved
 for $g=0$, but is not for $g\ne 0$. The conserved
one for $g\ne 0$ is 
\begin{eqnarray}
{\mathcal J}\equiv {\mathcal L}+\frac{1}{2}\tau^y,
\end{eqnarray}
 which satisfies $[{\mathcal H}_{\rm int}, \mathcal J]=0$. 
We find that eigenvalues of ${\mathcal L}$ are all integers, since
$
{\tilde S}^y=\sum_{n\ge 0}{\tilde S}_n^y
$ with
${\tilde S}_n^y=\frac{1}{2}{\mathcal L_n}$
 constituting the SU(2) algebra together with the
partners ${\tilde S}^{i}=\sum_{n\ge 0}{\tilde S}_n^{i}$ $(i=x,z)$, where
${\tilde S}^{i}_n$'s are given as
\begin{eqnarray}
{\tilde S}_n^z&\equiv&\frac{1}{4n}[{\mathcal
Q}_n{\mathcal Q}_{-n}-{\mathcal T}_n{\mathcal T}_{-n}+(n\to -n)],\\
{\tilde S}^x_n&\equiv&\frac{1}{2n}({\mathcal Q}_n {\mathcal T}_{-n}+{\mathcal
T}_n{\mathcal Q}_{-n}), 
\end{eqnarray} 
 for $n>0$, and 
\begin{eqnarray}
{\tilde S}_0^z&\equiv& \frac{1}{4}({\mathcal Q}_0^2-{\mathcal
T}_0^2+\theta_{\mathcal Q}^2-\theta_{\mathcal T}^2),\\ 
{\tilde S}_0^x&\equiv& \frac{1}{2}({\mathcal
 Q}_0{\mathcal T}_0+\theta_{\mathcal Q}\theta_{\mathcal T}).
\end{eqnarray}
${\tilde S}_n^{i}$'s satisfy $[{\tilde S}_n^i,{\tilde S}_m^j]=i\epsilon^{ijk}\delta_{nm}{\tilde S}_n^k$, and this leads to the
SU(2) algebra of $[{\tilde S}^i,{\tilde S}^j]=i\epsilon^{ijk}{\tilde S}^k$.
Thus, the eigenvalues of ${\mathcal L}$ are all integer,
and $\mathcal L$ is, indeed, the ``orbital'' angular momentum.
From this, the eigenvalues of $\mathcal J$ should be all half integers. 
Noticing that the total Hamiltonian is invariant under $\mathcal J$$\to$
$-\mathcal J$, {\it e.g.,} $\Uparrow \leftrightarrow \Downarrow$ and
${\mathcal Q}$$\to$$-\mathcal Q$, all the energy eigenstates should be
degenerate, which is the direct consequence of the unscreened state. 

\subsection{Fermi liquid phase} For $T_K\gg J$, the situation is 
conventional and the ground state is a FL. 
Each $d$-electron spin forms a spin singlet with 
conduction electrons. This phase is continuously connected to the
ground state of the spin $S=2$ four-channel Kondo problem. The way of  
fully screened processes are similar
to those in a Kondo singlet phase in the two-impurity Kondo model, where it is
essentially an $S=1$ two-channel model.\cite{2imp} Since there is
an asymmetry between $A_1$ and $T_2$ orbitals in Eq. (\ref{H2}), screening
takes place in two stages as $T$ decreases. Indeed, we have confirmed
the Fermi liquid properties by using Monte Carlo simulation as is  
explained in the next section.



\section{Continuous-time quantum Monte Carlo analysis} \label{Monte}
Now, we unveil the NFL properties and determine the global phase
diagram of this model by using CTQMC.\cite{CTQMC} 
 All the data presented
below are for $U$$=$$-2\epsilon$$=$$1.5D$, $t$$=$$0.2D$, and $k_Fa$$=$$3$. With these
values, $J$ is dominated by the superexchange $\sim$$4t^2/U$
with a small antiferromagnetic contribution of 
 the  RKKY term. For other values of $k_Fa$, the results are qualitatively the
 same as long as $J$ is antiferromagnetic.


\begin{figure}[th!]
\begin{center}
    \includegraphics[width=0.45\textwidth]{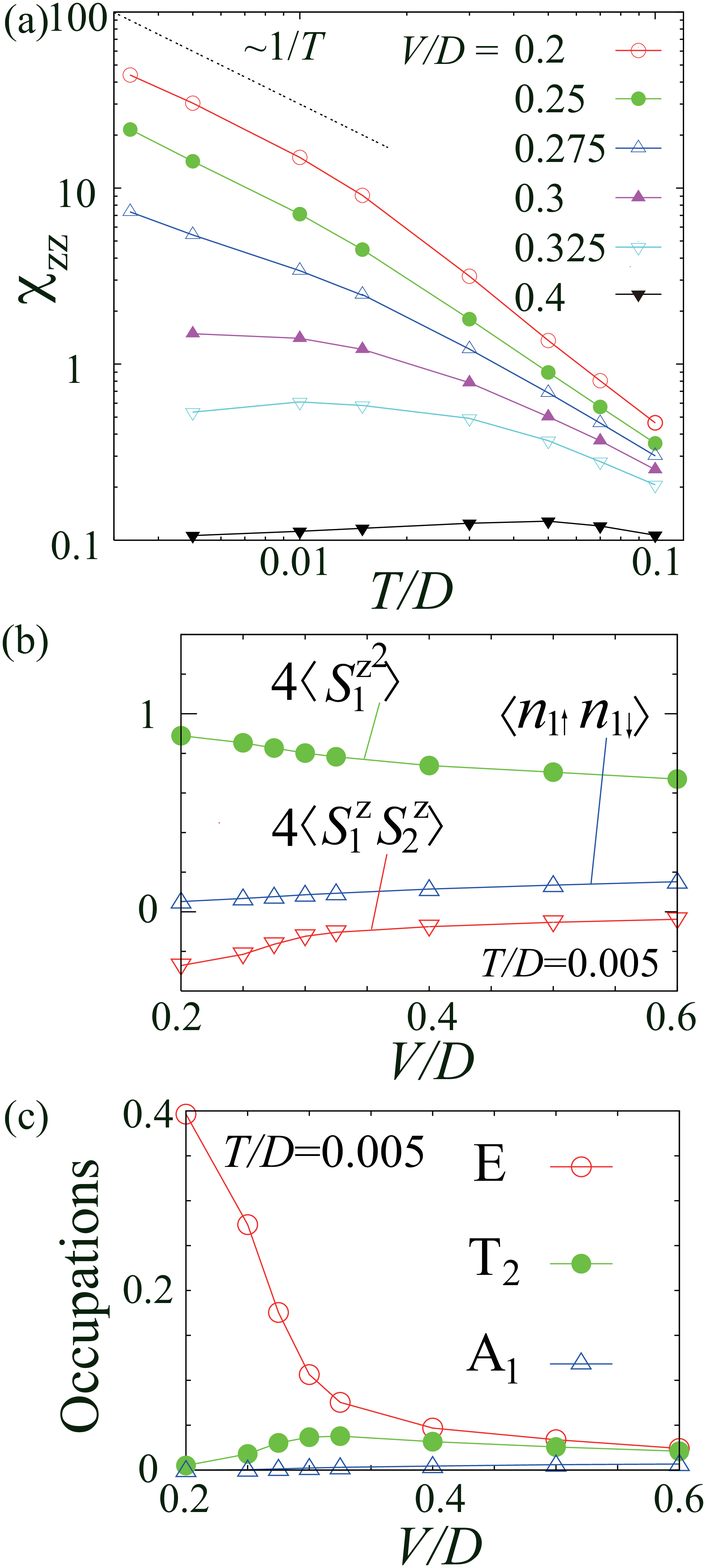}
\end{center}
\caption{(Color online). (a) $T$ dependence of 
 susceptibilities $\chi_{\rm zz}$ for ${V}/D=0.2-0.4$,
 $U=-2\epsilon=1.5D$, and $t=0.2D$. (b) Equal-time correlation
 functions: $\langle {S_1^z}^2\rangle$, $\langle
 n_{1\uparrow}n_{1\downarrow}\rangle$, and $\langle S_1^zS_2^z
 \rangle$. (c) ${V}$ dependence of the occupancies of
the  $E$, $T_2$ and $A_1$ states (per spin and orbital) in the four-electron sector. }
\label{fig-3}
\end{figure}
\subsection{Thermodynamics}
Figure {\ref{fig-3}}(a) shows  the susceptibility of local pseudospin
$\chi_{\rm zz}$ for
several values of ${V}$. 
 For ${V}\simle 0.275D$, 
 the chirality exhibits unscreened behaviors 
$\chi_{\rm zz}$$\sim$$T^{-1}$ at low $T$, 
which is consistent with the discussions
above. 
For larger ${V}$,
$T_K \gg J$, and thus the ground state is a FL.
 There, $\chi_{\rm zz}$ is small and constant at low $T$. 
Figure \ref{fig-3}(b) shows that the double occupancy $\langle
n_{1\uparrow}n_{1\downarrow}\rangle\sim 1/2-2\langle {S_1^z}^2\rangle$ increases 
 as ${V}$ increases. 
This implies that the $d$ electrons become more itinerant as
${V}$ increases. The intersite spin correlation $\langle
S_{1}^zS_2^z\rangle$ is large and 
antiferromagnetic for small ${V}$, while it is suppressed for large ${V}$.
As for local configurations in 
the four-electron sector, the occupancy of the 
ground-state doublet $E$ with spin $S=0$ 
decreases with increasing ${V}$, 
while those for the
first and the second excited states $T_2$($S=1$) and $A_1$($S=2$) 
increase as shown in Fig.{\ref{fig-3}}(c). 
This is a natural consequence of the Kondo screening, which 
mixes local configurations with different $S$.

\subsection{Single-electron dynamics}
\begin{figure}[t]
\begin{center}
    \includegraphics[width=0.45\textwidth]{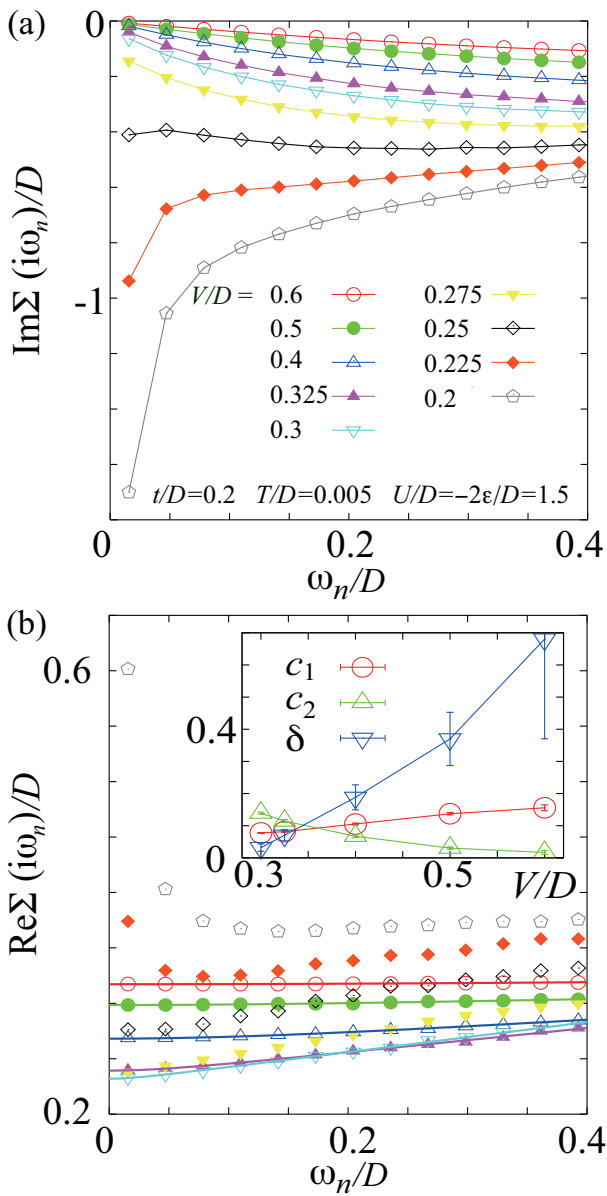}
\end{center}
\caption{(Color online). $\omega_n$ dependences of
 (a) Im$\Sigma(i\omega_n)$ and (b) Re$\Sigma(i\omega_n)$ for $T=0.005D$.  
 Lines for ${V}\ge 0.3D$ in panel (b) represent fit by
 $c_1+c_2\sqrt{\delta^2+\omega_n^2}$ and the values of $c_1$, $c_2$, and
 $\delta$ are shown in the inset.}
\label{fig-4}
\end{figure}

A drastic change also appears  in the single-electron dynamics upon varying
${ V}$. Figure
\ref{fig-4}(a) shows the imaginary part of the electron self-energy 
$\Sigma(i\omega_n)$ for the impurity $T_2$ orbital, for  $T=0.005D$ as a function of the
Matsubara frequency $\omega_n$. 
For large ${V}> {V}^*\simeq 0.3D$,
Im$\Sigma(i\omega_n)$ is linear in $\omega_n$ 
for small $\omega_n$,  as expected for the FL state. 
For smaller ${V}$, it shows a diverging behavior instead
 and this   
indicates that the $T_2$ electrons are localized in
the unscreened SC state. This is consistent with our conclusion based on
the exchange model (\ref{HK}), since the localization of the $T_2$ electron
 leads to the absence of resonance peak at the Fermi level in
the conduction-electron Green's function for 
the exchange model (\ref{HK}). 
The crossover around ${V}^*$ 
 is similar to that near a critical end point of
metal-insulator transitions in strongly correlated systems, {\it e.g.,}  
Mott's transition.\cite{Mott}


The real part Re$\Sigma(i\omega_n)$
shows an even more peculiar behavior near ${V}^*$ as shown in
Fig.\ref{fig-4}(b). Re$\Sigma(i\omega_n)$ for $V\sim V^*$ shows a
nonanalytic behavior, $\sim c_1+c_2|\omega_n|$, where
$c_1$ and $c_2$ are constants.
 The  $\omega_n$ dependence for ${V}\ge {V}^*$ is well fitted by 
  $\sim c_1+c_2\sqrt{\delta^2+\omega_n^2}$ with $\delta$$>$$0$. 
As shown in the inset, the critical point is realized by $\delta\to 0$. 
This $\omega_n$ dependence leads to the NFL
 form of the 
 retarded self-energy: 
\begin{eqnarray}
-{\rm Im}\Sigma^R(\omega)\sim|\omega|, 
\end{eqnarray}
which  is the same as in the marginal FL
 theory.\cite{MFL}
The NFL point ${V}^*\sim 0.3D$ is consistent with the transition point 
 of ${V}$ in Fig.\ref{fig-3}.
Thus, our CTQMC results for finite temperatures 
strongly suggest that there is a quantum critical point separating the FL 
and the unscreened phases at $T=0$.

As for the $A_1$ electron, the imaginary part of the self-energy shows
FL behavior 
for all the values of ${V}$ we examined. This persists for smaller
$t$ and is related to the fact that the
 $A_1$ orbital cannot interact with the nonmagnetic $E$ doublet apart from
 potential scattering.

This NFL is stable against various perturbations which are relevant at
some other types of critical points. For example, 
it is stable against magnetic field and
particle-hole asymmetry, and this is understood from the difference in 
the ground-state entropy in the two stable phases. Distortion that breaks
 the $T_d$ symmetry may give nontrivial effects on the stability of the
NFL. Examination of the stability of the NFL against distortions and
clarifying the NFL finite-size spectra  are our future work.


\section{Summary} \label{sum} We have investigated an Anderson model with four impurities 
on a regular tetrahedron. We have found that the system has two stable fixed
points. One is the new fixed point where an emerged SC 
 of the tetrahedron is not screened by
conduction electrons for small $V$ with diverging effective mass and 
the emergent PF excitations, and is a new class of singular FL states.
 The other is a Kondo screened
FL state for large ${V}$.
 Our CTQMC results have revealed
that the new critical NFL state appears, accompanying the $d$-electron
localization transition at zero temperature and  the
 self-energy exhibits marginal FL-like frequency dependence.

 
\section*{Acknowledgment}
The authors thank H. Kusunose and  J. Otsuki for their advice on Monte Carlo codes. 
This work is supported by KAKENHI (Grants No. 19052003 and No. 30456199) and 
by the Next Generation Super Computing Project, Nanoscience Program, from the MEXT
of Japan. A part of the numerical calculations was done at the
Supercomputer Center at the ISSP, University of Tokyo.


\end{document}